\documentclass[a4paper,fleqn,usenatbib]{mnras}

\usepackage[T1]{fontenc}
\usepackage{ae,aecompl}

\usepackage{graphicx}	
\usepackage{amsmath}	
\usepackage{amssymb}	

\usepackage{array}
\usepackage{rotfloat}
\usepackage{multirow}
\usepackage{esint}

\providecommand{\tabularnewline}{\\}

\title[Formation of LMXBs from wide binaries]{Tidal capture formation of Low Mass X-Ray Binaries from wide binaries in the field}

\author[E.\ Michaely and H.\ B.\ Perets]{Erez Michaely and Hagai B. Perets
\\
Physics Department, Technion -- Israel Institute of Technology, Haifa
3200004, Israel}

\date{Accepted XXX. Received YYY; in original form ZZZ}

\pubyear{2015}

\begin{document}

\label{firstpage}
\pagerange{\pageref{firstpage}--\pageref{lastpage}}
\maketitle

\begin{abstract}
We present a potentially efficient dynamical formation scenario for
Low Mass X-ray Binaries (LMXBs) in the field, focusing on black-hole
(BH) LMXBs. In this formation channel LMXBs are formed from wide binaries
($>1000$ AU) with a BH component and a stellar companion. The wide
binary is perturbed by fly-by's of field stars and its orbit random-walks
and changes over time. This diffusion process can drive the binary
into a sufficiently eccentric orbit such that the binary components
tidally interact at peri-center and the binary evolves to become a
short period binary, which eventually evolves into an LMXB. The formation
rate of LMXBs through this channel mostly depends on the number of
such BH wide binaries progenitors, which in turn depends on the velocity
kicks imparted to BHs (or NSs) at birth. We consider several models
for the formation and survival of such wide binaries, and calculate
the LMXB formation rates for each model. We find that models where
BHs form through direct collapse with no/little natal kicks can give
rise to high formation rates comparable with those inferred from observations.
This formation scenario had several observational signatures: (1)
the number density of LMXBs generally follows the background stellar
density and (2) the mass function of the BH stellar companion should
be comparable to the mass function of the background stellar population,
likely peaking at $0.4-0.6$ M$_{\odot}$. The latter aspect, in particular,
is unique to this model compared with previously suggested LMXB formation
models following common envelope binary stellar evolution. We note
that NS LMXBs can similarly form from wide binaries, but their formation
rate through this channel is likely significantly smaller due to their
much higher natal kicks. 
\end{abstract}


\section{Introduction}

Low Mass X-ray Binaries (LMXBs) are binaries composed of a compact
object (CO; Neutron star; NS; or a black hole; BH) accreting from
a lower mass, typically main sequence (MS) stellar companion with
$M_{*}<2M_{\odot}$. The CO accretes through an accretion disc that
generates X-ray emission (see \citet{PortegiesZwart1997,Tauris2006,Li2015}
for overviews). Most LMXBs are found in the field, but $\sim10\%$
\citep{Irwin2005} are found in globular cluster (GC). Given that
only $\sim0.1\%$ of stars reside in GCs, this result suggests that
the environment of GCs is highly conducive to the formation of LMXBs.
It is thought that close encounters in GCs give rise to tidal capture
of stars by COs, and the high formation rate of LMXBs. However, the
origin of field LMXBs is still debated, as binary stellar evolution
models encounter various challenges in producing LMXBs, and in particular
formation of BH LMXBs \citep{Eggleton1986,Kalogera1996,Kalogera1998,Kalogera1998a,Ivanova2006,Ivanova2010a},
where the outcomes of the common envelope phase pose major difficulties
\citep[e.g. ][and references therein]{Li2015,Wiktorowicz2014}. 

The distribution of LMXBs is not well understood.\citet{Voss2007}
studied the radial distribution of LMXBs in the bulge of M31. They
found that the radial density distribution in the \textit{bulge} of
M31 is consistent with a power law $n_{\rm LMXBs}\propto\rho_{*}^{\beta}$
and $\beta=2$, but the density profile becomes more shallow towards
the galaxy outer regions. For field LMXBs at larger distances it appears
that the number of LMXBs scale like the stellar light, which translates
to direct scaling with the stellar number density \citep[e.g. ][]{Gilfanov2004,Fabbiano2006,Paolillo2011}.
This suggests that the LMXBs systems are formed in-situ, namely in
the field and not in dense environments like GC. 

Most studies of field LMXBs explored their formation through binary
stellar evolution of isolated close binaries \citep{Eggleton1986,Kalogera1996,Kalogera1998,Kalogera1998a,Ivanova2010a}.
\citet{Voss2007} suggested they may form through tidal capture processes
similar to those occurring in GCs, at least in the inner dense region
of galaxies; this, however, may not apply to most LMXBs outside the
inner dense regions of galaxies. Here we suggest a novel scenario
in which field LMXB could originate from very wide binary progenitors
(semi-major axis;$a>1000$ AU), each composed of a CO and a low mass
MS star. Such wide binaries can be perturbed by encounters with field
stars and be excited into highly eccentric orbits. A sufficiently
high eccentricity can then give rise to a close peri-center approach
of the companion star of the CO, resulting in a tidal interaction
and the formation of a compact binary LMXB, somewhat similar to the
tidal capture formation of LMXBs in GCs. This approach follows the
recent study by \citet{Kaib2014} who showed that the components of
wide main sequence stellar binaries in the field could have high rates
of collisions/mergers through such a process. 

Motivated by this approach we calculated the formation rates of LMXBs
in the field from such processes. We find that under plausible conditions
this formation channel could explain the population of field LMXBs,
and in particular BH LMXBs, as well as give rise to unique predictions
about their properties. 

This paper is organized as follows. In section \ref{sec:Formation-of-LMXBs}
we present the theory for the formation of a LMXB from a wide binary.
In section \ref{sec:Channels-of-Formation} we specify the main channels
of formation, and present the expected formation rate in the Galactic
disk. In section \ref{sec:Discussion} we discuss the results and
summarize.

\section{Formation of LMXBs From Wide Binaries}
\label{sec:Formation-of-LMXBs}

\subsection{Basic Formation Scenario }

Let us consider a wide binary, $a>10^{3}AU$, with a stellar mass
BH or NS primary and a MS stellar companion. The system is affected
by two different perturbations. First, the galactic tide by the host
galaxy and in our case is the Milky Way (MW). Second, short duration
dynamical interactions with field stars. Here we focus on the latter,
likely stronger effect. The dynamical encounters can typically be
modeled through the impulse approximation, i.e. in the regime where
the interaction time is much shorter than the orbital period time.
These perturbations can torque the system, change the orbital angular
momentum, and exchange orbital energy thereby decreasing/increasing
the binary semi-major axis $a$. If such effects drive the system
to a sufficiently small periastron passage, $q$, tidal effects on
the MS star become important. Such tides can potentially produce significant
dissipation even during a single close approach and drive the binary
into a short period orbit, as suggested to occur in tidal capture
scenarios in dense stellar systems. The small separation between the
CO and the MS can then lead to a Roche lobe over flow (RLOF) onto
the compact object, and consequently produce a LMXB.

\subsection{Analytic Description }

In this section we derive the formation rate of LMXBs from wide, $a>10^{3}AU$,
binary systems. The derivation follows the same approach as \citet{Kaib2014}
and \citet{Hills1981}, however, here we consider the applications
and implications for the formation of LMXBs; we refer the reader to
these papers for additional details. Here we briefly review the calculation,
highlighting the main points and the minor differences arising from
the consideration of non equal-mass binaries. 

Consider an ensemble of isotropic wide binaries, each consisting of
a MS star with mass $m_{*}$ and radius $R_{*}$ and a CO with mass
$m_{\rm CO}$. The binaries are assumed to have the same SMA, with a thermal
distribution of eccentricities ($f(e)de=2ede$). We derive the probability
of forming LMXBs from this ensemble and find its dependence on the
SMA of the binaries, $a$. \citet{Kaib2014} showed that for MS-MS
wide binary, tidal circularization becomes important for pericenter
approaches below the tidal barrier $q_{\rm T}$, $q_{\rm T}\approx5R_{*}$.
For our CO-MS case we need to correct this tidal barrier. First, we
assume that no tidal forces act on the CO star (which radius is negligible
in this context), which decreases the tidal dissipation effect by
a factor of 1/2. Second, we need to account for the change in the
tidal force due to the change of $m_{*}\rightarrow m_{\rm CO}$, this
gives rise to additional a factor of 
\begin{equation}
\alpha\equiv m_{\rm CO}/m_{*}.\label{eq:alpha_define}
\end{equation}
Combining the two factors gives: 
\begin{equation}
q_{T}\approx5R_{*}\cdot\frac{1}{2}\cdot\frac{m_{\rm CO}}{m_{*}}=\frac{5R_{*}}{2}\alpha.\label{eq:q_T}
\end{equation}

Orbits which periastron becomes equal to or smaller than $q_{T}$
will circularize/tidally disrupt or collide with the CO and no longer
be wide binaries; such orbits are termed loss cone orbits. The fraction
of orbits occupying the lose cone for thermalized ensemble is 
\begin{equation}
F_{\rm q}=\frac{5R_{*}}{a}\alpha.\label{eq:F_q}
\end{equation}

After one orbital period, all wide binaries with loss-cone orbits
are ``destroyed'' (in the sense that they are no longer wide binaries)
as they approach peri-center. Other binaries outside the loss-cone
can be perturbed as to change their angular momentum and replenish
the loss cone. The average size of the phase-space region into which
stars are perturbed during a single orbital period is termed the smear
cone, defined by 
\begin{equation}
\theta=\frac{\left\langle \Delta v\right\rangle }{v_{b}}\label{eq:theta-1}
\end{equation}
 where $v_{b}$ is the binary MS companion velocity for a distance
of $1.5a$ with SMA of $a$, and $\left\langle \Delta v\right\rangle $
is the average change in the velocity over an orbital period due to
perturbations \citep{Hills1981}. Let us consider fly-by interactions
using the impulse approximation, where the interaction is considered
to be in the impulsive regime for which the interaction time, $t_{\rm int}$,
is much shorter than the binary period, $t_{\rm int}\ll P$. Hills shows
that on average the velocity change (for a binary with SMA, $a$),
to the binary components is of the order of 
\begin{equation}
\left\langle \Delta v\right\rangle \simeq\frac{3Gam_{*}}{v_{*}b_{*}^{2}}\label{eq:Delta_V}
\end{equation}
where $v_{*}$ is the velocity of the fly-by star with respect to
the binary center of mass and $b_{*}$ is the closest approach distance
of a fly-by (for the complete derivation see \citealp{Hills1981}).
In Kaib and Raymond's calculation $v_{b}=\left(G\mu/3a\right)^{1/2}$
where $\mu$ is the reduced mass of the binary. In our case the binary
is composed of two different mass components and we get 
\begin{equation}
v_{b}=\left(\frac{G\mu\alpha}{3a}\right)^{1/2}.\label{eq:v_binary}
\end{equation}
Hence the angular size of the smear cone cause by the impulse of the
fly-by on the binary is 
\begin{equation}
\theta=\frac{3Gam_{*}}{v_{*}b_{*}^{2}}\left(\frac{3a}{G\mu\alpha}\right)^{1/2}=9\frac{m_{*}}{\mu\alpha}\frac{v_{b}}{v_{*}}\left(\frac{a}{b_{*}}\right)^{2}\label{eq:theta_calc}
\end{equation}
and for $\theta\ll1$ we get the fractional size of the smear-cone
over the $4\pi$ sphere to be 
\begin{equation}
F_{s}=\frac{\pi\theta^{2}}{4\pi}=\frac{27}{4}\left(\frac{m_{*}}{\mu\alpha}\right)^{2}\left(\frac{G\mu\alpha}{av_{*}^{2}}\right)\left(\frac{a}{b_{*}}\right)^{4}\label{eq:F_s}
\end{equation}
The only difference between our expression and the one presented in
\citet{Kaib2014} and \citet{Hills1981} is a factor of $1/\alpha$. 

The condition for the loss cone to be continuously full is that the
sizes of the lose cone and the smear cone to be equal, i.e. the loss-cone
orbits are replenished at least as fast as they are depleted due to
the tidal interactions at peri-center. This equilibrium occurs when
\begin{equation}
\frac{F_{s}}{F_{q}}=\frac{1}{\alpha^{2}}\frac{27}{20}\left(\frac{m_{*}}{\mu}\right)^{2}\left(\frac{G\mu}{R_{*}v_{*}^{2}}\right)\left(\frac{a}{b_{*}}\right)^{4}=1.\label{eq:F_s/F_q}
\end{equation}

In this case, the full loss cone case rate of binaries being ``lost''
from our ensemble depends on the size of the loss cone (i.e. the fraction
of binaries are on lost-cone orbits), and on the period time of the
binary itself, the time it takes for a binary with a loss-cone orbit
to be destroyed 
\begin{equation}
\dot{L}_{q}=\frac{F_{q}}{P}=F_{q}\left(\frac{GM_{b}}{4\pi^{2}a^{3}}\right)^{1/2}\label{eq:Loss_rate}
\end{equation}
where $M_{b}=m_{\rm CO}+m_{*}$ is the mass of the binary and $P$ is
the period of the binary. Note that the loss rate is independent of
the stellar density in the field, i.e. once the stellar density is
sufficiently large as to fill the loss-cone, the loss rate is saturated,
and becomes independent of the perturbation rate. Furthermore, one
can see from Eq. (\ref{eq:Loss_rate}) that the full loss-cone rate
scales like $\dot{L}\propto a^{-3/2}$, i.e. the full loss-cone rate
decreases with increasing SMA. As one goes to smaller SMAs, the loss
rate increases until it approaches $F_{\rm s}/F_{\rm q}\rightarrow1$, at
which point the loss-cone is no longer full. At this regime, the empty
loss cone regime, the rate is not determined by the size of the loss
cone, but by the rate in which binaries are perturbed into the loss-cone.
Hence in order to calculate the loss rate in the empty loss cone regime
we need to estimate the rate of encounters. We can express the rate
of flyby encounters by 
\begin{equation}
f=n_{*}\sigma v,\label{eq:f_encounter_Rate}
\end{equation}
where $n_{*}$ is the stellar density, $\sigma=\pi b_{*}^{2}$ is
the geometric cross-section and $v$ is the velocity of the fly-by's.
Given the stellar density, $n_{*}$ we need to find an expression
for the geometric cross-section and the velocity. Eq. (\ref{eq:F_s/F_q})
gives us the condition for the crossover from the full to empty loss
cone regime. Equating Eq. (\ref{eq:F_s/F_q}) to unity we find the
condition 
\begin{equation}
\left(v_{*}b_{*}^{2}\right)^{2}\leq\frac{1}{\alpha^{2}}\frac{27}{20}\left(\frac{m_{*}}{\mu}\right)^{2}\left(\frac{G\mu a^{4}}{R_{*}}\right),
\end{equation}
and the the rate $f$ is given by 
\begin{equation}
f=n_{*}\pi\frac{1}{\alpha}\frac{m_{*}}{\mu}\sqrt{\frac{27}{20}\frac{G\mu a^{4}}{R_{*}}}.\label{eq:f_value}
\end{equation}
Note that the loss rate of binaries increases with $f$ only until
it reaches the orbital frequency. Beyond this point we are in the
full loss-cone regime and the loss rate saturates and remains constant. 

Combining these considerations together, one can see that at small
separations the loss-rate is determined by the empty loss-cone and
increases with the binary SMA, until we reach a critical SMA, $a_{\rm crit}$,
at which the loss cone becomes full, and the loss rate is determined
by the full loss-cone, and now decreases with larger SMAs. The maximal
rate is therefore obtained for binaries with the critical SMA. To
compute $a_{\rm crit}$ we equate $f$ with the orbital frequency, $f=\left(GM_{b}/4\pi^{2}a^{3}\right)^{1/2}$
and solve for $a_{crit}$: 
\begin{equation}
a_{\rm crit}=\left[\frac{5}{27}\frac{\alpha^{2}}{\pi^{4}}\frac{\mu M_{b}}{m_{*}^{2}}\frac{R_{*}}{n_{*}^{2}}\right]^{1/7}\label{eq:a_crit}
\end{equation}
which differs from \citet{Kaib2014} by a factor of $2\alpha^{2}$,
i.e. for binaries with more massive compact object, the empty loss
cone regime becomes more important as the $a_{\rm crit}$ value is larger. 

The loss probability is calculated differently for the two different
regimes. For the empty loss cone regime the limiting factor is the
value of the function $f$. $F_{q}$ is the fraction of wide binaries
destroyed, and therefore $\left(1-F_{q}\right)$ represents the fraction
of binaries that survive as wide binaries at the relevant timescale.
For the full loss cone regime the relevant timescale is $1/f$; for
the empty loss cone regime the relevant timescale is $P$. Therefore,
this term is a monotonically decreasing function of time, and the
probability for a wide binary to closely interact at peri-center and
no longer survive as a wide binary is 
\begin{equation}
L_{a<a_{\rm crit}}=1-\left(1-F_{\rm q}\right)^{t\cdot f}\label{eq:empty_Prob}
\end{equation}
where $t$ is the time since birth of the binary. As one can expect
the probability only depends on the size of the loss cone and the
rate of interactions. For the limit of $t\cdot f\cdot F_{\rm q}\ll1$
we can expand eq. (\ref{eq:empty_Prob}) and take the leading term,
to find the loss probability to be approximated by 
\begin{equation}
L_{a<a_{crit}}=t\cdot f\cdot F_{q}.\label{eq:empty_prop_approx}
\end{equation}
Substituting the function $f$ from eq. (\ref{eq:f_value}) and the
value $F_{q}$ from eq.(\ref{eq:F_q}) we then get 
\begin{equation}
L_{a<a_{\rm crit}}=t\cdot n_{*}\cdot m_{*}\cdot a\sqrt{\frac{135\pi^{2}}{4}\frac{GR_{*}}{\mu}.}\label{eq:empty_propb_express}
\end{equation}
Note that there is no $\alpha$ dependence in this case, and the probability
differs from that obtained by \citet{Kaib2014} only by a numerical
factor.

For the full loss cone regime the limiting factor is not the value
of $f$, but rather the orbital period and/or the orbital frequency.
Therefore, the full expression for the loss probability for $a>a_{\rm crit}$
is
\begin{equation}
L_{a>a_{\rm crit}}=1-\left(1-F_{\rm q}\right)^{t/P}.\label{eq:full_prob}
\end{equation}
 For the limit $F_{q}\cdot t/P\ll1$ we can approximate the probability
by 
\begin{equation}
L_{a>a_{crit}}=t\cdot F_{q}\cdot\frac{1}{P}=t\cdot\alpha\cdot\sqrt{\frac{25GM_{b}R_{*}^{2}}{4\pi^{2}a^{5}},}\label{eq:full_prob_apx_express}
\end{equation}
where the probability is increased by a factor of $\alpha$ in this
case. 

The above calculations provide the estimated rates for a simplified
scenario, where binaries are randomly perturbed and continue to evolve
until they produce LMXBs when their components have close interactions.
In reality, the perturbations may also ``ionize'' a binary and destroy
it, namely, the binary is disrupted by the random fly-by's. Such ionization
process decreases the available number of wide binaries, and consequently
lowers the rates of LMXBs formation. To account for the ionization
process we consider the finite lifetime of wide binaries due to flybys
using the approximate relation given by \citet{Bahcall1985} for $t_{1/2}$,
the half-life time of a wide binary evolving through encounters 
\begin{equation}
t_{\rm 1/2}=0.00233\frac{v_{*}}{Gm_{*}n_{*}a}.
\end{equation}
Taking this into account we can correct for eq. (\ref{eq:empty_propb_express})
and eq. (\ref{eq:full_prob_apx_express}) to get 
\begin{equation}
L_{a<a_{\rm crit}}=\tau\cdot n_{*}\cdot m_{*}\cdot a\sqrt{\frac{135\pi^{2}}{4}\frac{GR_{*}}{\mu}}\left(1-e^{-t/\tau}\right)\label{eq:ionization empty}
\end{equation}
and 
\begin{equation}
L_{a>a_{\rm crit}}=\tau\cdot\alpha\cdot\sqrt{\frac{25GM_{b}R_{*}^{2}}{4\pi^{2}a^{5}}}\left(1-e^{-t/\tau}\right),\label{eq:ionization full}
\end{equation}
where $\tau=t_{\rm 1/2}/\ln2$ is the mean lifetime of the binary.

\subsubsection{LMXB formation rate dependence on stellar density\label{sub:Dependence-on-stellar}}

Using the above model to calculate the probability for close interactions
leading to LMXB formation we can now derive an overall estimate for
the formation rate of LMXBs, by considering realistic distributions
for the initial conditions for the wide binaries population and stellar
number densities. In order to do so we need to integrate the loss
probabilities weighted by the SMA distribution of wide binaries for
all SMA in the binaries ensemble. We assume that the SMA distribution
of wide binaries follows a log-uniform distribution in $a$, namely
$f_{\rm a}\sim1/a$, where $f_{\rm a}$ is the SMA distribution of the wide
binaries population \citep{Duchene2013}. We consider binaries in
the range from $a_{\rm min}=10^{3}AU$ up to $a_{\rm max}=3\cdot10^{4}AU$,
where the maximal value is due to cutoff by the Galactic tide. 

This estimated rate mostly depends on $a_{\rm crit}$. The values of $a_{\rm crit}$
can be divided to two distinct regimes: the empty cone which scales
like the stellar density $n_{*}$ and the full cone regimes which
is constant in $n_{*}$. For the expression of $a_{\rm crit}$ in eq.
(\ref{eq:a_crit}) we find that there exists a value of $n_{*}=n_{*0}$
for which $a_{\rm crit}=a_{max}$. This implies that for a given stellar
density satisfying $n_{*}<n_{*0}$ the ensemble is always in the empty
loss cone regime and therefore the probability scales like $n_{*}$
and the number of formed LMXB systems scales like $n_{*}^{2}$. This
argument neglects the ionization process described earlier. Binary
ionization changes the scaling. Following (\ref{eq:ionization empty}),
we can write the equation explicitly to get
\[
L_{\rm empty}=\frac{0.00233\cdot v_{*}}{Gm_{*}n_{*}a}\cdot n_{*}\cdot m_{*}\cdot a\sqrt{\frac{135\pi^{2}}{4}\frac{GR_{*}}{\mu}}\left(1-e^{-t/\tau}\right)
\]
 
\begin{equation}
L_{\rm empty}\propto\sqrt{\frac{135\pi^{2}}{4}\frac{GR_{*}}{\mu}}\left(1-e^{-t/\tau}\right),
\end{equation}
which has no density dependence. In our case it should be noted that
typically $t\gg\tau$ for the wide binaries we consider. We therefore
conclude that once ionization is accounted for, the formation probability
of LMXBs in this regime is density independent. To obtain the total
formation rates we therefore only need to multiply the probability
for LMXB formation per system by the total number of systems. In other
words, the total number of formed system depends only linearly on
the number density of wide binaries, which follows the background
stellar density; hence, the number density of formed LMXB systems
scales like the stellar number density, $n_{*}$ in this regime. 

For densities larger than $n_{*0}$, which comprise only a small part
of the galaxy, we have contributions from both the empty loss cone
and the full cone regimes. If we consider the dependence on density
in this regime we need to add the different components to get
\[
P_{\rm LMXB}=\int_{a_{\rm min}}^{a_{\rm crit}}L_{a<a_{\rm crit}}f_{\rm a}da\cdot dN\left(r\right)+
\]
\[
\int_{a_{\rm crit}}^{a_{\rm max}}L_{a>a_{\rm crit}}f_{\rm a}da\cdot dN\left(r\right)=
\]
 
\begin{equation}
C_{\rm 1}\cdot n_{*}\cdot\ln n_{*}+C_{\rm 2}n_{*},\label{eq:density_power}
\end{equation}
where $C_{\rm 1}$ and $C_{\rm 2}$ are some calculable constant pre-factors.
Note that the density dependence differs for different $f_{\rm a}$ distributions. 

\begin{figure*}
\includegraphics[width=\columnwidth]{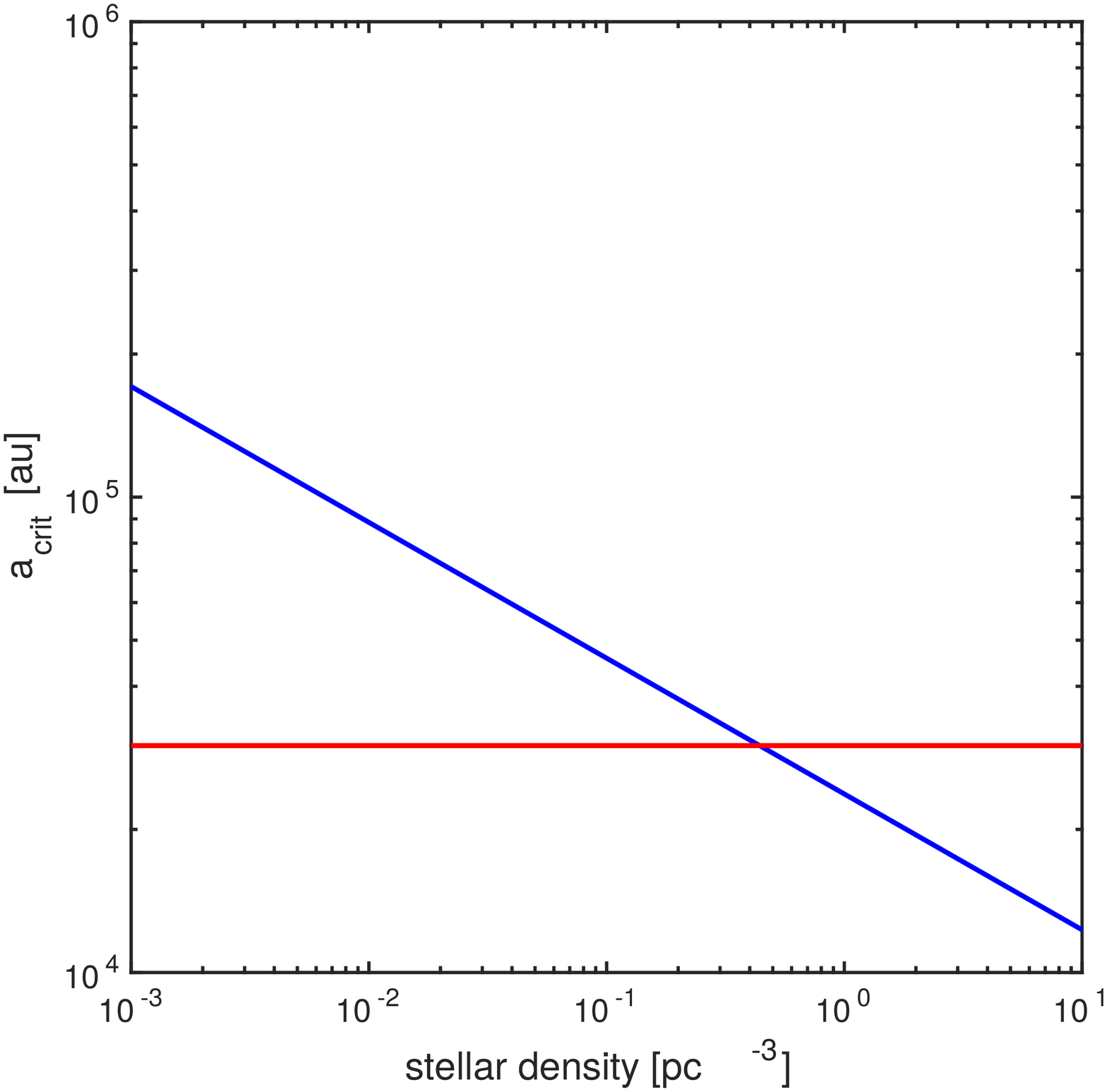}\hspace{0.5cm}
\includegraphics[width=\columnwidth]{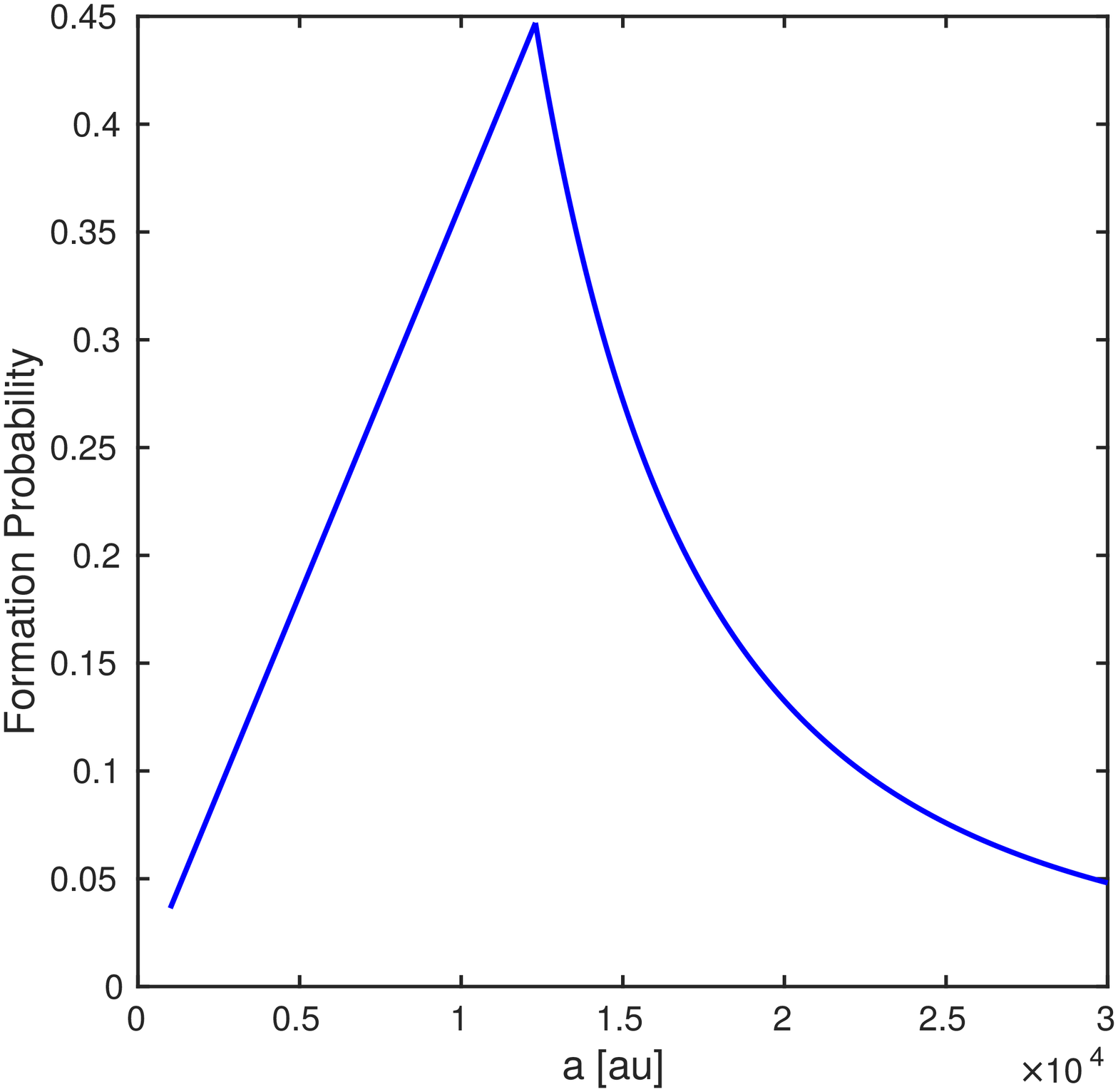}

\protect\caption{\label{fig:a_crit vs density-1} The empty and full loss-cone regions
as a function of the stellar density and binary separation. Left:
$a_{\rm crit}$ as a function of the stellar density $n_{*}$ (considering
binaries with a primary mass of $m_{CO}=10M_{\odot}$, and a secondary
with $m_{*}=0.4M_{\odot}$and $R_{*}=0.4R_{\odot}$. The red solid
horizontal line marks the value of $a_{\rm max}=3\cdot10^{4}AU$. Once
$a_{\rm crit}$ becomes larger than than $a_{\rm max}$ all binaries in the
ensemble are in the empty loss-cone regime, and the loss cone rate
depends linearly on $n_{\star}$. For $a_{\rm crit}<a_{\rm max}$ there are
contributions from both the empty and full loss cone regimes. Right:
Loss-cone rate dependence on the binaries separation (for the case
of $a_{\rm crit}<a_{\rm max})$. Shorter period binaries are in the empty
loss-cone regime, while wider binaries $a_{\rm crit}<a<a_{\rm max}$ are in
the full loss-cone regime; the loss cone rate dependence on the SMA
differs between these regimes, and the maximal rate is obtained for
$a_{\rm crit}$(see text). The dependence on the stellar density in this
regime therefore depends on the specific contribution of binaries
in the full and the empty loss cone regimes (see Eq. \ref{eq:density_power}),
which in turn depends on the relative contribution from binaries with
$a>a_{\rm crit}$and those with $a<a_{\rm crit}$}
\end{figure*}

To illustrate these issues, let us consider an ensemble of binaries
all of which with a primary CO component with $m_{CO}=10M_{\odot}$,
a secondary of $0.4{\rm M_{\odot}}$ and , an age of $t=10Gyr,$ and
the secondary radius of $R_{*}=0.4R_{\odot}$ (which affects the the
strength of the tidal interactions). In Fig. \ref{fig:a_crit vs density-1}
(left panel) we show the relation between $a_{crit}$ and $n_{*}$
for the values described above (blue solid line). The red dashed line
indicates the value of $a_{\rm max}$ and the intersection indicates the
value of $n_{*0}$. In Fig. \ref{fig:a_crit vs density-1} (right
panel) we see the formation probability of LMXBs for $n_{*}>n_{*0}$
and the two regimes are noticeable. The linear slope corresponds to
eq. \ref{eq:empty_propb_express} and the decaying curve corresponds
to eq. \ref{eq:full_prob_apx_express} (see very similar fig. in \citealp{Kaib2014}). 

\subsubsection{Calculation of the total number of LMXBs}

In order to calculate the total number of field LMXBs formed in the
Galactic disk, we need to integrate over the contribution from all
regions of the Galaxy. Let us then consider $dN\left(r\right)=n_{*}\left(r\right)\cdot2\pi\cdot r\cdot h\cdot dr$
to be the the number of stars in a region $dr$ (and scale height
$h$), located at distance $r$ from the center of the galaxy. Following
\citet{Kaib2014} and references within we model the Galactic stellar
density in the Galactic disk as follows 
\begin{equation}
n_{*}\left(r\right)=n_{0}e^{\left(-\left(r-r_{\odot}\right)/R_{l}\right)},
\end{equation}
where $n_{0}=0.1pc^{-3}$ is the stellar density near our sun, $R_{l}=2.6kpc$
\citep{Juric2008} is the galactic length scale and $r_{\odot}=8kpc$
is the distance of the sun from the galactic center. Integrating over
the stellar densities throughout the galaxy we can obtain the total
yield of LMXBs in the Galaxy through this process. The results of
these calculations are discussed in depth in the next sections.

\subsubsection{Caveats, assumptions and uncertainties}

Before continuing to discuss various channels for LMXB formation through
wide binaries scatterings (and the number of LMXBs they produce),
we briefly mention several important caveats, uncertainties and explicit
assumptions used in the suggested model. 
\begin{itemize}
\item We tacitly assume that all tidal captures of a MS star by a CO leads
to the formation of an LMXB, as similarly assumed by previous authors
discussing tidal capture formation of LMXBs in dense stellar environments.
However, the evolution of a tidally captured binary may not always
lead to LMXB formation and other end products may also result from
such evolution. For a detailed discussion of the post-capture evolution
of binaries see \citet{Ray1987}
\item The SMA distribution of wide binaries, and in particular high mass
binaries is not well determined. In eq. \ref{eq:density_power} we
use the value of $f_{\rm a}$ to be log uniform, as suggested by limited
observational data for low mass wide binaries, and suggested theoretical
studies for the formation of wide binaries \citep{Duchene2013}.
\item The actual lifetime of LMXBs is not well determined; here we use a
typical lifetime discussed in the literature \citep{Verbunt1993}
of 1 Gyr. 
\end{itemize}

\section{LMXB formation channels}

\label{sec:Channels-of-Formation}In this section we estimate the
number of LMXBs formed by wide binary progenitors in the Galaxy, based
on the detailed calculations described in the previous sections (see
section \ref{sec:Formation-of-LMXBs}). The formation rates strongly
depend on the frequency of the wide massive binary progenitors. Since
the latter is not yet determined observationally, we consider various
possible formation channels of wide-binaries containing a NS or a
BH, and consider the implications of the different channels on the
formation rates of LMXBs.

In subsections \ref{sub:Single-BH} and \ref{sub:Single-NS-via} we
describe the formation of BH-MS and NS-MS wide binaries, respectively,
through the cluster-dispersal scenario in which a single compact object
captures a wide stellar companion following the dispersal of the host
cluster; in subsection \ref{sub:Binaries} we consider primordially
formed wide binaries which survive the stellar evolution stage leading
to the birth of the compact object. The resulting wide CO-binary populations
in each of these scenarios also depend on specific assumptions regarding
the formation of the compact objects and their natal kicks, if such
occur. Tables \ref{tab:Number-of-singleBH-MS}-\ref{tab:BH-MS binary}
summarize the different formation channels, the various assumptions
made for each specific model, and the total number of LMXBs expected
to have formed through these scenarios.

\subsection{Cluster-dispersal -- capture-formed wide binary progenitors }

\label{sub:Single-BH}Most stars are born in stellar clusters and
associations, which later disperse \citep{Lada2003}. Following the
cluster dispersal, stars may leave the cluster with small relative
velocities and become bound binaries \citep[e.g. ][]{Kouwenhoven2010,Moeckel2010,Perets2012a}.
Such post-dispersal formed binaries typically have very wide orbits
which could explain the origin of wide binaries in the field. In the
following we consider this scenario for the origin of wide binaries
and its implications for the formation of LMXBs.

\subsubsection{BH formation with a natal kick}

Consider a single BH which still resides in its host open cluster
after its formation. \citet{Perets2012a} showed that as an open cluster
dissolves massive objects are more likely to capture a companion star
and form wide binaries in the cluster-dispersal scenario. Primaries
of mass $m_{\rm primary}>5M_{\odot}$ have a wide binary fraction (BF)
of $f_{\rm BF}\approx1/2$ . A fraction of $f_{\rm WF}\approx0.6$ of these
wide binaries have separations in the range $10^{3}AU<a<3\cdot10^{4}AU$,
and we can therefore find the number of wide binaries formed through
this process. However, we still need to determine how many BHs still
reside in their host clusters after their violent birth. In particular,
if BHs are kicked upon formation they may escape their host cluster
and will not be able to dynamically capture a wide companion.

The existence and the amplitude of BH natal kicks is still not understood
(see \citealp{Repetto2015} for a recent overview), and we therefore
consider several possible cases. 

\begin{figure*}
\includegraphics[width=\columnwidth]{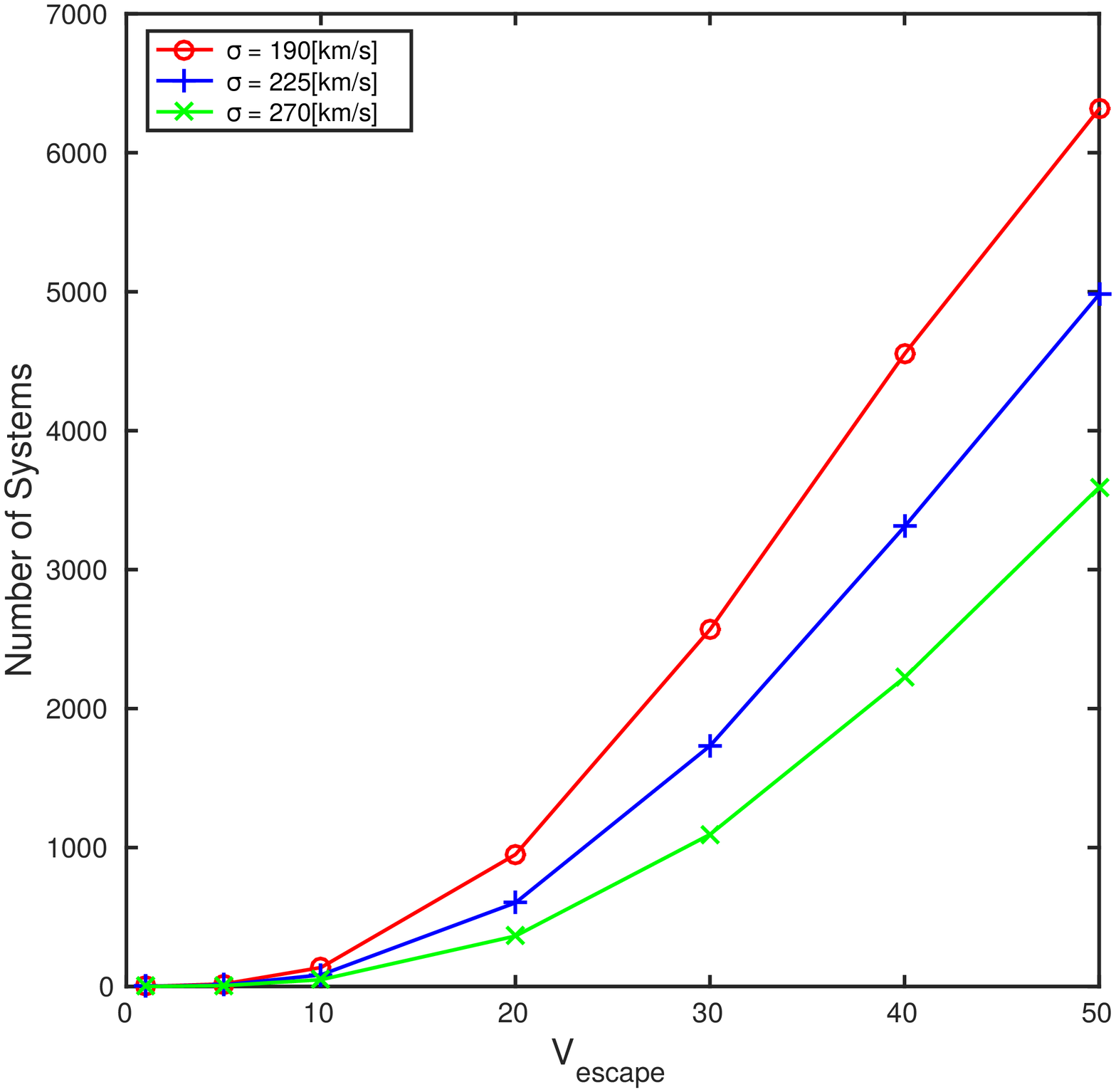}\hspace*{0.5cm}
\includegraphics[width=\columnwidth]{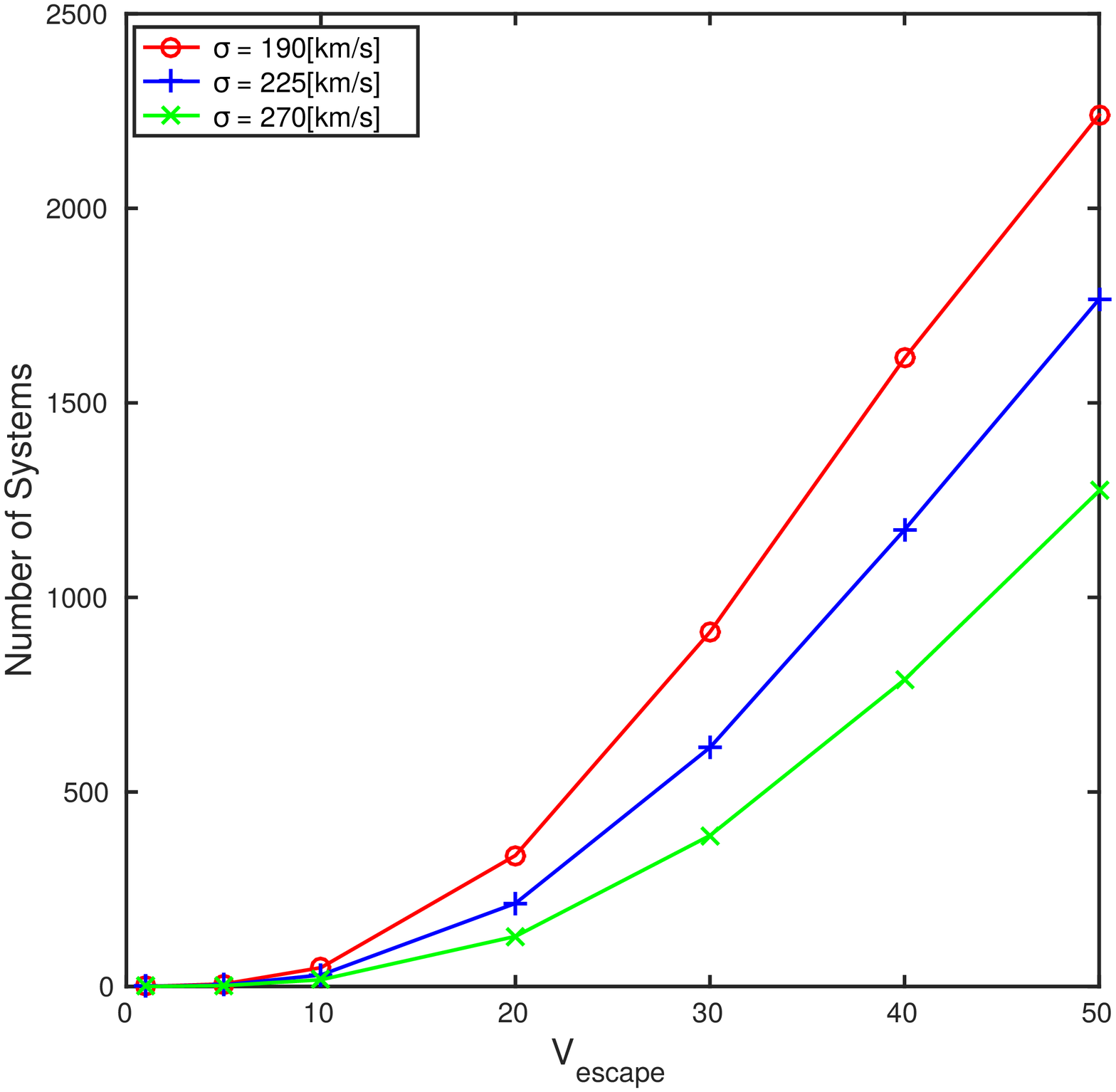}
\protect\caption{\label{fig:Single_BH_kick}The number of LMXB systems formed through
the single BH + cluster-dispersal capture channel as a function of
the cluster escape velocity. All BHs in this calculation receive momentum
kicks, using Maxwellians velocity distribution for NSs, with several
possible velocity dispersions, consistent with observations of young
pulsars velocities. Red circles correspond to the case of $\sigma=190kms^{-1}$
; blue pluses to $\sigma=225kms^{-1}$ ; and green X's correspond
to $\sigma=270kms^{-1}$. The BH masses are taken to be $\left\langle M_{\rm BH}\right\rangle =10M_{\odot}$.
The left and right panels correspond to the LMXB numbers obtained,
without and with binary ionization, respectively. }
\end{figure*}

We first consider the case in which BH kicks are derived from the
natal kicks of NSs that form in the process, before collapsing to
become BHs. In this case we use a Maxwellian distribution for the
velocities of NS natal kicks and explore three different velocity
dispersions $\sigma_{\rm NS}=190,\ 225,\ 270kms^{-1}$\citet{Fryer2001,Arzoumanian2002},
consistent with observations of the velocity distribution of young
pulsars. We first assume that the BHs have the same \textit{momentum}
kick as NSs, and therefore their actual kick velocities are inversely
proportional to the ratio between the mass of the BH and the mass
of a typical NS, namely $\left\langle m_{\rm BH}\right\rangle /\left\langle m_{\rm NS}\right\rangle $
\citep{Fryer2001}. For simplicity we generally use an average BH
mass of $\left\langle M_{\rm BH}\right\rangle =10M_{\odot}$. We calculate
analytically the fraction of the BH which have a velocity kick smaller
than the escape velocity of the cluster $v_{\rm esc},$ where we consider
several values for the cluster escape velocity $v_{\rm esc}=1,5,10,20,30,40,50kms^{-1}$.
We can now find the number of non-escaping BHs, and thereby find the
total number of wide BH-MS binaries formed from dissolving star clusters.
Finally, we can use these to derive $N_{\rm LMXB}$, the number of LMXBs
formed from such binaries
\[
N_{\rm BH-LMXB}=\int P_{\rm LMXB}da\cdot dN\times f_{O}\times f_{\rm low-kick}\times
\]
 
\begin{equation}
f_{\rm Single}\times f_{\rm BF}\times f_{\rm WF}\times\frac{t_{\rm life}}{t},
\end{equation}
where $\int P_{\rm LMXB}dadN$ is the formation rate calculation described
in the previous sections, where we use typical values for the BH and
MS stars: $m_{*}=0.4M_{\odot}$, $M_{\rm BH}=10M_{\odot}$, $R_{*}=0.4R_{\odot}$
and taking $t=10Gyr$ for calculating the overall time for wide binary
scattering; $f_{O}\sim0.001$ is the fraction of O-star progenitors
of BHs (assuming a Salpeter initial mass function) ; $f_{\rm Single}\sim0.2$
is the fraction of single stars among O-stars (binaries will be considered
later on) \citep{Duchene2013}. $f_{\rm BF}\sim0.5$ is the wide binary
fraction as described above; $f_{\rm WF}\sim0.6$ is the wide binary fraction
in the relevant separation range, as mentioned above. $f_{\rm low-kick}$
is the fraction of BHs formed with sufficiently low velocity kick
as to be retained in the cluster before its dispersal. In order to
turn the formation rates into observed numbers, one needs to account
for the lifetime of the LMXB, which is assumed to be $t_{\rm life}\sim10^{9}yr$.
Taking these together we obtain the expected number of LMXBs in the
galaxy. Fig. \ref{fig:Single_BH_kick} shows the number of LMXB systems
created through this channel with the parameters mentioned above.
In Fig. \ref{fig:Single_BH_kick} (left plot) we present the numbers
for the three different velocity dispersions $\sigma=190,\ 225,\ 270kms^{-1}$,
without accounting for the binary ionization. Fig. \ref{fig:Single_BH_kick}
(right plot) shows the number of LMXBs after accounting for binary
ionization.

\begin{table*}
\begin{tabular}{|>{\centering}m{3cm}|>{\centering}p{4cm}|c|c|c|c|c|c|c|}

\hline 
\multirow{2}{3cm}{Formation Channel} & \multirow{2}{4cm}{Assumptions} & \multirow{2}{*}{$v_{\rm escape}$ {[km/s]}} & \multicolumn{2}{c|}{$\sigma=190km/s$} & \multicolumn{2}{c|}{$\sigma=225km/s$} & \multicolumn{2}{c|}{$\sigma=270km/s$}\tabularnewline
\cline{4-9} 
 &  &  & $N_{\rm no\ ion.}$ & $N_{\rm ion.}$ & $N_{\rm no\ ion.}$ & $N_{\rm ion.}$ & $N_{\rm no\ ion.}$ & $N_{\rm ion.}$\tabularnewline
\hline 
\hline 
\multirow{4}{3cm}{Single BH + capture} & \multirow{4}{4cm}{momentum kick for all progenitor masses. $\left\langle m_{\rm BH}\right\rangle =10M_{\odot}$;
$f_{\rm BF}=0.5$} & $1$ & $0.14$ & $0.05$ & $0.08$ & $0.03$ & $0.05$ & $0.017$\tabularnewline
\cline{3-9} 
 &  & $5$ & $17.6$ & $6.25$ & $10.65$ & $3.7787$ & $6.18$ & $2.2$\tabularnewline
\cline{3-9} 
 &  & $10$ & $136$ & $49$ & $83$ & $29.5$ & $48.6$ & $17.2$\tabularnewline
\cline{3-9} 
 &  & $20$ & $950$ & $337$ & $602$ & $214$ & $363$ & $129$\tabularnewline
\hline 

\end{tabular}

\vspace{0.5cm}

\begin{tabular}{|>{\centering}m{3cm}|>{\centering}p{6cm}|>{\centering}m{2.5cm}|>{\centering}p{2cm}|c|}
\hline 
Formation Channel & Assumptions & $v_{\rm escape}$ & $N_{\rm no\ ion.}$ & $N_{\rm ion.}$\tabularnewline
\hline 
\hline 
Single BH + capture & no momentum kick for $M_{\rm progenitor}>30M_{\odot}$; $\left\langle m_{\rm BH}\right\rangle =10M_{\odot}$;
$f_{\rm BF}=0.5$ & don't depend on $v_{\rm esc.}$ & $3100$ & $1100$\tabularnewline
\hline 
\end{tabular}

\caption{\label{tab:Number-of-singleBH-MS}The number of LMXB systems formed
through the single BH + cluster-dispersal capture channel. The BH
mass is assumed to be $\left\langle m_{\rm BH}\right\rangle =10M_{\odot}$.
The upper part of the table shows the results for different natal
kicks distributions (see text and Fig. \ref{fig:Single_BH_kick}).
Results are shown for several possible type of clusters with escape
velocities ranging between 1--50~km~s$^{-1}$. The lower part
of the table shows the results for the case where no natal kick is
imparted to the BHs, for the cases where BHs with massive progenitors
($M_{\rm progenitor}>30M_{\odot}$) go through direct collapse and do
not receive a natal kick. Binary ionization significantly decreases
the number of available progenitors and hence the resulting number
of LMXBs (compared with the cases where it is not accounted for).}
\end{table*}

\subsubsection{BH formation through direct collapse without a natal kick}

\label{sub:singe-BH-without-kick}If we assume that BH progenitors
with MS mass greater than $30M_{\odot}$ form a BH through a direct
collapse without a SN \citep{Belczynski2004,Belczynski2006}, the
numbers of formed LMXB are significantly larger, as the host cluster
retains many more BHs; effectively all of the BHs formed by stellar
progenitors with $m_{progenitor}>30M_{\odot}$, irrespective of the
cluster escape velocity. Note that the the BH mass is not correlated
with the progenitor mass in this range, and therefore this formation
channels should not bias the BH masses in the formed LMXBs . We find
the total number of formed BH field LMXBs (with and without accounting
for ionization) to be
\[
N_{\rm LMXB}=\int P_{\rm LMXB}da\cdot dN\times f_{\rm Q\ BH}\times f_{\rm Single}\times
\]
 
\begin{equation}
f_{\rm BF}\times f_{\rm WF}\times\frac{t_{\rm life}}{t}\approx3100,
\end{equation}
\[
N_{\rm ion-LMXB}=\int P_{\rm LMXB-ion.}da\cdot dN\times f_{\rm Q\ BH}\times f_{\rm Single}\times
\]
\begin{equation}
f_{\rm BF}\times f_{\rm WF}\times\frac{t_{\rm life}}{t}\approx1100.
\end{equation}

\subsubsection{Single NS formation via electron capture SN}

\label{sub:Single-NS-via}
\begin{figure}
\includegraphics[width=\columnwidth]{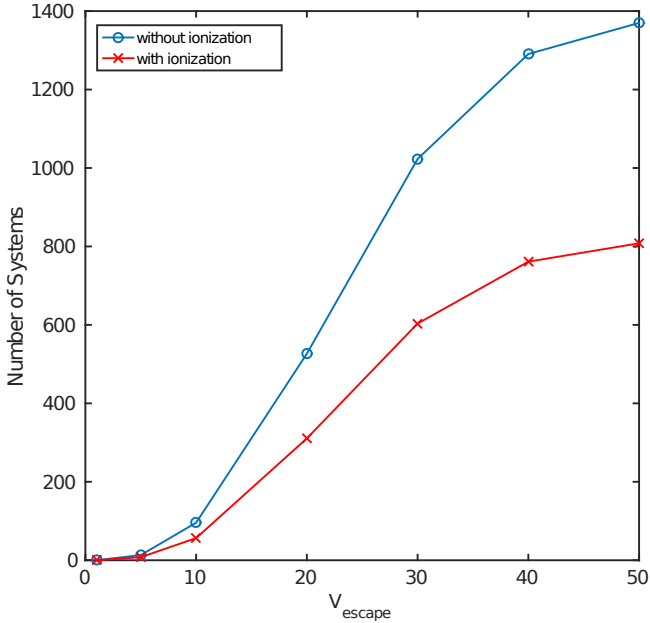}

\protect\caption{\label{fig:electron-capture-SN}The number of NS-LMXB systems formed
from single electron-capture NS + cluster-dispersal channel as a function
of the cluster escape velocity. The NS mass taken to be $\left\langle M_{\rm NS}\right\rangle =1.33M_{\odot}$
and the Maxwellian kick velocity dispersion is $\sigma=15\left[km/s\right]$.
blue upper line (circles): the number of systems without calculating
ionization from fly-by's; red lower line (crosses): the number of
systems with calculating ionization from fly-by's.}
\end{figure}

The typical NS natal kicks based on the
young pulsar velocity distribution \citep{Fryer2001} are much higher
than the escape velocity of any stellar cluster in the Galaxy, and
therefore a negligible number of NSs are expected to be retained in
clusters and to capture a wide companion following the cluster dispersal.
However, a subsample of NSs is known to have much lower velocity dispersion,
and are thought to possibly form through an electron capture SN process
\citep{Nomoto1984,Nomoto1987}. In electron capture (EC) SN the natal
kick of the newly born NS get is much lower than in the regular core
collapse case. We model the natal kick values with a Maxwellian distribution
with a velocity dispersion of $\sigma_{\rm EC}=15\left[km\cdot s^{-1}\right]$.
For this value of $\sigma_{EC}$ $99\%$ of the kicks are lower than
$50\left[km\cdot s^{-1}\right]$. In this channel we consider every
single O star in the mass range $8M_{\odot}<M_{\rm progenitor}<10M_{\odot}$
to become a NS through electron capture SN, and we assume the NS mass
to be $\left\langle M_{\rm NS}\right\rangle =1.33M_{\odot}$ and use the
same type of calculation as before to get
\[
N_{\rm NS-LMXB}=\int P_{\rm LMXB-ion.}da\cdot dN\times f_{\rm NS}\times f_{\rm Low\ Kick}\times
\]
 
\begin{equation}
f_{\rm Single}\times f_{\rm BF}\times f_{\rm WF}\times\frac{t_{\rm life}}{t},
\end{equation}
where $f_{\rm NS}=7\cdot10^{-4}$ is the fraction of electron capture
NS produced assuming a Salpeter IMF and the appropriate progenitor
mass range; $f_{\rm Single}=0.2$ is the fraction of single O star progenitors
(binaries are treated below); $f_{\rm BF}=0.2$ and $f_{\rm WF}=0.6$ are
taken from \citet{Perets2012a}. In Fig. \ref{fig:electron-capture-SN}
we present the number of NS-LMXB systems produced via this mechanism
as a function of the escape velocity of the cluster (see Table \ref{tab:Electron capture NS}
and Fig. \ref{fig:electron-capture-SN}). 

\begin{table*}
\begin{tabular}{|>{\centering}p{3.5cm}|>{\centering}m{5cm}|c|>{\centering}m{2cm}|>{\centering}m{2cm}|}
\hline 
Formation Channel & Assumptions & $v_{\rm escape}[km/s]$ & $N_{\rm no\ ion.}$ & $N_{\rm ion.}$\tabularnewline
\hline 
\hline 
\multirow{4}{3.5cm}{Single NS + capture} & \multirow{4}{5cm}{electron capture SN for $8M_{\odot}<M_{\rm progenitor}<10M_{\odot}$;
$\sigma=15km/s$; $f_{\rm BF}=0.2$} & $1$ & $0.1$ & $0.06$\tabularnewline
\cline{3-5} 
 &  & $5$ & $13.2$ & $7.78$\tabularnewline
\cline{3-5} 
 &  & $10$ & $95.7$ & $56.5$\tabularnewline
\cline{3-5} 
 &  & $20$ & $527$ & $310$\tabularnewline
\hline 
\end{tabular}\protect\caption{\label{tab:Electron capture NS}The number of NS-LMXB systems formed
from single electron-capture formed NS + cluster-dispersal channel
as a function of the cluster escape velocity. The NS mass is taken
to be $\left\langle M_{\rm NS}\right\rangle =1.33M_{\odot}$.}
\end{table*}

\subsubsection{Capture formed wide triples}

Previously we considered captured formed binaries. In principle non-wide
binaries can also capture an additional wide orbit third companion
via the same mechanism described above. This newly formed hierarchical
triple system can similarly be perturbed in the field by fly-by's.
However, as the orbit of the wide companion becomes eccentric, the
triple system will be driven into an unstable configuration resulting
in a strong chaotic interaction typically followed by ejection of
one of the triple components binary. An unstable configuration occurs
when the peri-center of the third companion is comparable to a few
times the SMA of the primordial binary. During the chaotic evolution,
a stellar component could potentially flyby close to the CO and be
tidally captured. However, the fraction of systems from the overall
phase space that eventually achieve such a close approach is (following
\citealp{Valtonen2006}) 
\begin{equation}
f\left(\left|E_{B}\right|\right)d\left|E_{B}\right|=3.5\left|E_{0}\right|^{7/2}\left|E_{B}\right|^{-9/2}d\left|E_{B}\right|
\end{equation}
where $E_{0}$ is the total energy of the triple system, and $E_{\rm B}$
is the energy of the surviving binary that could lead to a LMXB phase.
For LMXBs $f\approx10^{-5}$, hence the LMXB formation rate from this
channel is negligible. If the third companion is captured into a highly
inclined orbit with respect to the inner binary orbit, it could potentially
drive the inner binary into high eccentricities through Lidov-Kozai
secular evolution. This capture triggered secular evolution is beyond
the scope of this study and will be discussed elsewhere. Recently,
and independently, the Lidov-Kozai LMXB formation channel in triples
was discussed in details by \citet{Naoz2015}; earlier suggestions
of LMXBs formation through secular evolution in triples have been
explored by \citet{Mazeh1979}. Another origin of LMXB in destabilized
triples has been suggested by \citet{Perets2012}.

\subsection{Primordial binary progenitors}

\label{sub:Binaries}In the previous section we considered wide binary
progenitors formed through the cluster-dispersal capture scenario.
In the following we consider primordial wide binary progenitors.

\subsubsection{BH - MS wide binary without a natal kick\label{sub:wideBH--MS}}

Here we consider primordial wide binaries, where again we chose typical
values of $\left\langle M_{\rm BH}\right\rangle =10M_{\odot}$ and a $M_{\rm secondary}=0.4M_{\odot}$.
To calculate the frequency and distribution of such binaries, we have
used the BSE \citep{Hurley2002} binary evolution population synthesis
code where the initial binary population was created following the
observed binary properties as reviewed by \citet{Duchene2013}. We
find that a fraction of $f_{\rm w-BH-MS}\approx3\cdot10^{-5}$ of the
systems produce wide BH-MS binaries, assuming no natal kicks is given
to the BHs. The results are summarized in Table \ref{tab:BH-MS binary}.
Note that fraction of formed BH-MS wide binaries when momentum kick
is imparted to the BHs at birth is negligible (the escape velocity
of wide binaries is very low, and the kick easily disrupt them).

\begin{table*}
\begin{tabular}{|c|>{\centering}p{8cm}|c|c|}
\hline 
Formation Channel & Assumptions & $N_{\rm no\ ion.}$ & $N_{\rm ion.}$\tabularnewline
\hline 
\hline 
\multirow{2}{*}{Wide BH - MS binary} & momentum kick for all progenitor masses. $\left\langle m_{\rm BH}\right\rangle =10M_{\odot}$; $\sigma_{\rm BH}=19km/s$ & $165$ & $58$\tabularnewline
\cline{2-4} 
 & no momentum kick for $M_{\rm progenitor}>30M_{\odot}$; $\left\langle m_{\rm BH}\right\rangle =10M_{\odot}$ & $3440$ & $1220$\tabularnewline
\hline 
\end{tabular}\protect\caption{\label{tab:BH-MS binary}The number of systems created from the primordial
wide BH -MS binaries. The upper row: BH formation with natal momentum
kick with velocity dispersion of $\sigma_{\rm BH}=19km/s$. The bottom
row: BH formation with no natal kick.}
\end{table*}

\section{Discussion and Summary }

\label{sec:Discussion}In this work we explored a novel channel for
formation of BH-LMXBs in the field from wide binary progenitors. We
showed that a CO-MS wide binary ($>1000$ AU) that undergoes perturbations
from fly-by's in the field, can have a finite probability to be kicked
into an sufficiently eccentric orbit and be affected by tidal interactions.
It can then evolve into a short period orbit, and eventually produce
an X-ray binary once the stellar companion fills its Roch-lobe. We
find that the formation rate of LMXBs through this channel strongly
depends on the frequency of wide BH-MS binaries. The frequency of
such binaries is sensitive to the natal kick given to BHs at birth.
Silent formation of BHs from direct collapse with no natal kicks allows
for the existence of a significant number of wide BH-MS binaries that
can produce LMXBs. Depending on the formation scenario of the wide
binaries, either as primordial binaries or through the cluster-dispersal
capture scenario, we find that hundreds or thousands of BH-LMXBs could
have formed in the Galactic disk through this channel, respectively,
consistent with the numbers of field LMXBs inferred from observations.
However, if BHs receive natal kicks at birth with comparable momentum
kick as NS natal kicks, only a few $\sim10s-100s$ LMXBs could form
through this channel, depending on specific assumptions (see Table
\ref{tab:Number-of-singleBH-MS} for details). 

NS LMXBs form in negligible numbers for typical NS natal kicks. NSs
forming with low velocity natal kicks (e.g. possibly through electron-capture
channel) produce up to the formation of a few tens or more LMXBs,
depending on specific assumptions. 

The wide-binary progenitor model for the formation of LMXBs give rise
to specific observational signatures, which can differentiate them
from other models. In particular they suggest a natural origin for
the otherwise puzzling observed companion mass function, and the high
number of field LMXBs in low density environments. These can be summarized
as follows: 
\begin{itemize}
\item \textbf{Spatial distribution}: As discussed in Sec. \ref{sub:Dependence-on-stellar}
the distribution of LMXBs in the field scales with the background
stellar density like $C_{1}\cdot n_{*}\cdot\ln n_{*}+C_{2}n_{*}$;
where typically generally the distribution of LMXBs should follow
the stellar light \citep{Gilfanov2004,Fabbiano2006,Paolillo2011},
besides in the more dense regions in the inner parts of the Galaxy. 
\item \textbf{Companion mass function}: The BH-companion mass in observed
LMXBs is found to be in the range $0.1-1$ $M_{\odot}$, peaking at
0.6 $M_{\odot}$; this is inconsistent with all of the currently suggested
models for LMXBs formation through common-envelope evolution \citep{Wiktorowicz2014}.
However, in our suggested wide-binary progenitor scenario the mass
function of the BH companion in the LMXB is expected to follow the
mass function of stellar companions in massive wide binaries. In the
cluster-dispersal capture scenario origin of wide binaries, the companion
mass function should generally follow the mass function of field stars.
The observed mass ratio of O star binaries with wide companions scale
like $q^{-2}$ (priv. comm. Maxwell Moe 2015), where $q$ is the secondary
to primary mass ratio. This distribution is consistent with random
pairing with present day mass function, peaking at a few 0.1 $M_{\odot}$.
Our models therefore provides a unique signature and can explain the
companion mass function in LMXBs, which challenges other formation
scenarios. 
\item \textbf{Additional wide companions}: As mentioned here and explored
in depth by \citet{Naoz2015}, secular evolution in wide triples can
potentially provide another alternative model for LMXB formation without
a challenging common-envelope phase. One might then generally expect
the existence of a wide third companion to the LMXBs. Though not excluded,
such an additional companion is not required in the wide-binary progenitor
model. The frequency of third wide companions to LMXBs could therefore
potentially differentiate the two different models. 
\end{itemize}

\section*{Acknowledgements}

We acknowledge support from the I-CORE Program of the Planning and
Budgeting Committee and The Israel Science Foundation grant 1829/12.
We thank Tom Maccarone for stimulating discussions.

\bsp	
\label{lastpage}
\end{document}